\documentclass[aps,prd,onecolumn,groupedaddress,showpacs,nofootinbib,amssymb]{revtex4}
\usepackage[dvips]{graphicx}
\usepackage{amssymb}
\usepackage{amsmath}
\usepackage{graphicx,color}
\usepackage{amsfonts}
\usepackage{bm}
\usepackage{cancel}
\usepackage{comment}
\usepackage{hyperref}
\usepackage{ulem}

\newcommand{\e}{\mathrm{e}}

\allowdisplaybreaks[4]

\begin{document}

\tolerance=5000

\title{Extremely small stars in scalar-tensor gravity: when stellar radius is less than Schwarzschild one}
\author{Shin'ichi~Nojiri$^{1,2,3}$}\email{nojiri@gravity.phys.nagoya-u.ac.jp}
\author{Sergei~D.~Odintsov$^{3,4}$}\email{odintsov@ice.csic.es}
\author{Armen~Sedrakian$^{5,6}$}\email{sedrakian@fias.uni-frankfurt.de}

\affiliation{ $^{1)}$ Department of Physics, Nagoya University, Nagoya 464-8602, Japan \\
$^{2)}$ Kobayashi-Maskawa Institute for the Origin of Particles
and the Universe, Nagoya University, Nagoya 464-8602, Japan \\
$^{3)}$ Institute of Space Sciences (ICE, CSIC) C. Can Magrans
s/n, 08193 Barcelona, Spain \\
$^{4)}$ ICREA, Passeig Luis Companys, 23, 08010 Barcelona, Spain \\
$^{5)}$ Frankfurt Institute for Advanced Studies,
D-60438 Frankfurt am Main, Germany \\
$^{6)}$ Institute of Theoretical Physics, University
of Wroc\l{}aw, 50-204 Wroc\l{}aw, Poland 
}

\begin{abstract}
We show analytically that there exist compact stellar objects akin
to neutron stars whose radius is smaller than the Schwarzschild radius 
defined by Arnowitt-Deser-Misner (ADM) mass.
The radius of the compact object is defined by the radius where the
energy density and the pressure of ordinary matter vanish, while
clouds of scalar(s) can extend beyond this radius -- a situation
that is often encountered in modified gravity theories, like $F(R)$
gravity and the scalar--Einstein--Gauss-Bonnet gravity. 
The clouds of scalar mode(s) give additional contributions to the ADM mass and
as a result, the corresponding Schwarzschild radius given by the ADM mass can be larger 
than that of the compact object.
\end{abstract}


\maketitle

\section{Introduction}

In recent years, there has been a surge in the precision observations regarding compact objects, 
which delivered crucial constraints on the equation of state of stellar matter assuming that the underlying objects are compact stars. 
Notably, the most intriguing constraints include the pulsar PSR~J0740+6620~\citep{NANOGrav:2019jur, Fonseca:2021wxt}, 
setting a record with a neutron star's maximum mass of $2.0\, M_{\odot}$. 
There are also simultaneous mass and radius measurements for PSR~J0030+0451~\citep{Vinciguerra:2023qxq} and 
PSR~J0740+6620~\citep{Salmi:2022cgy}, extracted from X-ray observations of nearby pulsars. 
Finally, gravitational waves detected from the binary merger event GW170817 estimated 
the tidal deformability of the binary, resulting in constraining the radii of the individual components. 
A larger mass has been claimed for the companion of the ``black widow'' pulsar PSR~J0952-0607 
with the estimate of the mass $M > 2.19M_{\odot}$ (at $1\sigma$ confidence)~\citep{Romani:2022jhd}. 
Modelling the X-ray spectrum of the central compact object within the supernova remnant HESS~J1731-347 
combined with distance estimates from Gaia observations, indicating a very low mass of 
$M = 0.77^{+0.20}_{-0.17}\, M_{\odot}$ and a relatively small radius of $R = 10.4^{+0.86}_{-0.78}$~km 
(at $1\sigma$ confidence)~\citep{Doroshenko:2022}. 
Finally, one should mention the gravitational wave event GW190814 which involved 
a binary coalescence of a 24.3 $M_{\odot}$ black hole with a compact object 
in the mass the range of 2.50 -- 2.67~$M_{\odot}$ where the secondary might be either 
the heaviest observed neutron star or the lightest observed black hole. 

In the framework of General Relativity, one possible realisation of compact objects with very small radii is strange stars 
which arise from the Witten-Bodmer conjecture about the ground state of absolutely stable matter. 
Their radii can be much smaller than expected in ordinary neutron stars~\citep{Weber:2004kj}. 
The strange star hypothesis is compatible with current data on the mass, radius, 
and tidal deformability measurement from GW170817, but is one 
of the few possibilities that may furnish us with very compact stars. 
Another possibility for ultra-compact stars is the strong first-order phase transition 
to quark matter with an equation of state characterized by large speed of sound~\citep{Li:2022ivt}. 
However, the present conjecture may provide the only possible realization of highly massive neutron stars {\it and} have radii smaller than the Schwarzschild radius. 

Recently in \cite{Li:2023vbo}, it was claimed that within a modified
gravity a stellar object such as a neutron star may have a radius that
is smaller than the Schwarzschild radius defined by the subleading behavior of the metric 
for a large radius, which corresponds to the Arnowitt-Deser-Misner (ADM) mass~\cite{Arnowitt:1959ah}. 
The model in Ref.~\cite{Li:2023vbo} is called a quasi-topological gravity model and
includes the third power of the Ricci curvature and the scalar curvature. 
The paper by Li et al. \cite{Li:2023vbo} introduces a quasi-topological gravity model, 
incorporating the third power of the Ricci curvature and the scalar curvature. 
Typically, similar models exhibit a scalar mode present also in the standard $F(R)$ gravity
model, as well as a massive spin-two mode that manifests as a ghost. 
However, the model proposed in \cite{Li:2023vbo} is purportedly devoid of any ghosts, 
due to its quasi-topological properties.

While the findings presented in \cite{Li:2023vbo} are intriguing, they are not unexpected. 
Modified gravity models often introduce additional modes, like the scalar mode found in $F(R)$ gravity, 
which function akin to forms of matter. 
Consequently, when defining the radius of celestial bodies based solely on conventional matter, 
such as highly compressed solids present in the crusts of neutron stars, these modes 
create a sort of `hair' around the object or form a surrounding cloud. 
This complication makes it challenging to precisely delineate the boundary of this `hair'. 
Incorporating these modes into the radius definition despite the difficulty in pinpointing their
boundary could result in a radius larger than that of a black hole. 
Specifically, in the framework of $F(R)$ gravity, defining the radius solely by the boundary 
of compressed solid matter might yield a radius smaller than that of an ADM equivalent-mass black hole. 
This discrepancy could be due to the contribution of scalar hair and 
additional mass contributed by the cloud of the scalar mode. 

Numerous studies \cite{Capozziello:2015yza, AparicioResco:2016xcm, Astashenok:2017dpo, Astashenok:2018iav,
 Feola:2019zqg, Numajiri:2023uif, Nashed:2024pbc}, which involve careful and tedious numerical computations,
have studied the mass-radius relationship of neutron stars within the framework of $F(R)$ gravity~\cite{Nojiri:2010wj, Nojiri:2017ncd}.
Such models are then confronted with observations as for example,
 done in Ref.~\cite{Nashed:2024pbc} who show that the predictions of their $F(R)$ gravity model with
$F(R)=R\e^{\xi R}$ and constant $\xi$ are consistent with the observations of
Pulsar SAX J1748.9-2021 as well as other astrophysical constraints. 

In this paper, we show analytically that there exist gravity models
that admit a spherically symmetric solution for a stellar object with
a radius smaller than the Schwarzschild radius. For this purpose, we
consider Einstein's gravity coupled with two scalar fields
\cite{Nojiri:2020blr}. This model exhibits spherically symmetric
dynamical spacetimes, however, it commonly introduces ghosts~\cite{Kugo:1979gm}. 
To avoid these, it is possible to introduce constraints, employing
Lagrange multiplier fields \cite{Nojiri:2023dvf, Nojiri:2023zlp,
 Elizalde:2023rds}, which effectively eliminate these ghosts. These
constraints resemble those observed in the theoretical framework of
mimetic gravity~\cite{Chamseddine:2013kea}.

This paper is organised as follows. In Sec.~\ref{sec:Model}, we introduce our model of two scalars coupled 
to Einstein's gravity which is void of ghosts. 
The method by which the ghosts are eliminated from the theory is discussed in Sec.~\ref{sec:Elimitation}. 
In Sec.~\ref{sec:Reconstruction}, we find solutions in the case of general spherically symmetric and time-dependent spacetimes. 
The static limit is considered in Sec.~\ref{sec:StaticStars}. 
We discuss the formation of extremely compact objects in Sec.~\ref{sec:Formation}. 
Our results are summarised in Sec.~\ref{sec:Conclusions}. 

\section{Einstein's Gravity coupled to two scalars}
\label{sec:Model}

In Ref.~\cite{Nojiri:2020blr}, it has been shown that spherically symmetric spacetimes can be realized, 
even in the case where spacetime is dynamical, within Einstein's gravity that is coupled with two scalar fields. 
The model in \cite{Nojiri:2020blr}, however, contains ghosts. 
In subsequent work of Refs.~\cite{Nojiri:2023dvf, Nojiri:2023zlp, Elizalde:2023rds}, ghost-free models have been
proposed by introducing constraints via Lagrange multiplier fields. 
This approach bears similarity to the constraint that appears in the 
theoretical framework of mimetic gravity~\cite{Chamseddine:2013kea}.

We start with Einstein's gravity coupled with two scalar fields $\phi$ and $\chi$, whose action is given by, 
\begin{align}
\label{I8}
S_{\phi\chi} = \int d^4 x \sqrt{-g} & \left\{ \frac{R}{2\kappa^2}
 - \frac{1}{2} A (\phi,\chi) \partial_\mu \phi \partial^\mu \phi
 - B (\phi,\chi) \partial_\mu \phi \partial^\mu \chi \right. \nonumber \\
& \left. \qquad - \frac{1}{2} C (\phi,\chi) \partial_\mu \chi \partial^\mu \chi
 - V (\phi,\chi) + \mathcal{L}_\mathrm{matter} \right\}\, ,
\end{align}
where $V(\phi,\chi)$ is the potential for $\phi$ and $\chi$ fields and $\mathcal{L}_\mathrm{matter}$ 
is the matter Lagrangian. 
The variation of the action \eqref{I8} with respect to the metric $g_{\mu\nu}$ gives
\begin{align}
\label{gb4bD4}
0= &\, \frac{1}{2\kappa^2}\left(- R_{\mu\nu} + \frac{1}{2} g_{\mu\nu} R\right) \nonumber \\
&\, + \frac{1}{2} g_{\mu\nu} \left\{
 - \frac{1}{2} A (\phi,\chi) \partial_\rho \phi \partial^\rho \phi
 - B (\phi,\chi) \partial_\rho \phi \partial^\rho \chi
 - \frac{1}{2} C (\phi,\chi) \partial_\rho \chi \partial^\rho \chi - V (\phi,\chi)\right\} \nonumber \\
&\, + \frac{1}{2} \left\{ A (\phi,\chi) \partial_\mu \phi \partial_\nu \phi
+ B (\phi,\chi) \left( \partial_\mu \phi \partial_\nu \chi
+ \partial_\nu \phi \partial_\mu \chi \right)
+ C (\phi,\chi) \partial_\mu \chi \partial_\nu \chi \right\} 
+ \frac{1}{2} T_{\mathrm{matter}\, \mu\nu} \, ,
\end{align}
where $T_{\mathrm{matter}\, \mu\nu}$ is the energy-momentum tensor of matter. 
The field equations are obtained by varying the action \eqref{I8} with
respect to the fields $\phi$ and $\chi$ and are given by
\begin{subequations}
\label{I10} 
\begin{align}
0 =& \frac{1}{2} A_\phi \partial_\mu \phi \partial^\mu \phi
+ A \nabla^\mu \partial_\mu \phi + A_\chi \partial_\mu \phi \partial^\mu \chi
+ \left( B_\chi - \frac{1}{2} C_\phi \right)\partial_\mu \chi \partial^\mu \chi
+ B \nabla^\mu \partial_\mu \chi - V_\phi \, , \\
0 =& \left( - \frac{1}{2} A_\chi + B_\phi \right) \partial_\mu \phi \partial^\mu \phi
+ B \nabla^\mu \partial_\mu \phi
+ \frac{1}{2} C_\chi \partial_\mu \chi \partial^\mu \chi + C \nabla^\mu \partial_\mu \chi
+ C_\phi \partial_\mu \phi \partial^\mu \chi - V_\chi \, ,
\end{align}
\end{subequations}
where $A_\phi=\partial A(\phi,\chi)/\partial \phi$, etc. 
Note that these field equations are nothing but the Bianchi identities.

\section{Elimination of the ghosts}
\label{sec:Elimitation}

We now consider a general spherically symmetric and time-dependent
spacetime, whose metric is given by\footnote{For the argument
that the metric in (\ref{GBiv}) is a general one, see \cite{Nojiri:2020blr}.},
\begin{align}
\label{GBiv}
ds^2 = - \e^{2\nu (t,r)} dt^2 + \e^{2\lambda (t,r)} dr^2 + r^2 \left( d\vartheta^2 + \sin^2\vartheta d\varphi^2 \right)\, .
\end{align}
We also make the following identifications,
\begin{align}
\label{TSBH1}
\phi=t\, , \quad \chi=r\, .
\end{align}
This assumption does not lead to a loss of generality. For the
spherically symmetric solutions~(\ref{GBiv}) of the theory~(\ref{I8}),
$\phi$ and $\chi$ depend on both coordinates $t$ and $r$ in general.
Given a solution, the $t$- and $r$-dependence of $\phi$ and $\chi$ can
be determined, and $\phi$ and $\chi$ are then given by specific
functions $\phi(t,r)$, $\chi(t,r)$ of $t$ and $r$. We can now
redefine the scalar fields to replace $t$ and $r$ with new scalar
fields, say, $\bar{\phi}$ and $\bar{\chi}$ with
$\phi(t,r)\to \phi(\bar{\phi}, \bar{\chi})$ and
$\chi(t,r) \to \chi(\bar{\phi}, \bar{\chi})$. We can then identify
the new fields with $t$ and $r$ as in (\ref{TSBH1}),
$\bar{\phi}\to \phi=t$ and $\bar{\chi} \to \chi= r$. The change of
variables
$\left( \phi, \chi\right) \rightarrow \left( \bar{\phi} , \bar{\chi}
\right)$ can then be absorbed into the definitions of $A$, $B$, $C$,
and $V$ in the action~(\ref{I8}). Therefore, the
assumption~(\ref{TSBH1}) does not lead to loss of generality.
We should note, however, that there exists another class 
of solutions that do not satisfy Eq.~(\ref{TSBH1}), as discussed below.

Our next aim is to build a model capable of realizing spherically
symmetric geometry, given by the metric (\ref{GBiv}), which could be static or time-dependent. 
This, however, leads to instances where $A$ and/or $C$ assume negative values in the realized geometry. 
This indicates that $\phi$ and/or $\chi$ become ghosts. 
The presence of a ghost mode implies a negative kinetic energy in the classical theory 
and the generation of negative norm states in quantum theory~\cite{Kugo:1979gm}, 
rendering the model physically inconsistent. 

To address this issue, we proceed to eliminate these ghosts by 
imposing constraints on $\phi$ and $\chi$. 
This approach bears similarity to the mimetic constraint outlined in 
\cite{Chamseddine:2013kea}, effectively rendering the ghost modes non-dynamical or frozen. 
Thus, to eliminate the ghosts with the help of the Lagrange multiplier fields $\lambda_\phi$ and $\lambda_\chi$, 
we modify the action (\ref{I8})
$S_{\mathrm{GR} \phi\chi} \to S_{\mathrm{GR} \phi\chi} + S_\lambda$, 
where the additional term is given by
\begin{align}
\label{lambda1}
S_\lambda = \int d^4 x \sqrt{-g} \left[ \lambda_\phi \left( \e^{-2\nu(t=\phi, r=\chi)} \partial_\mu \phi \partial^\mu \phi + 1 \right)
+ \lambda_\chi \left( \e^{-2\lambda(t=\phi, r=\chi)} \partial_\mu \chi \partial^\mu \chi - 1 \right) \right] \, .
\end{align}
The variations of $S_\lambda$ with respect to $\lambda_\phi$ and
$\lambda_\chi$ provide the following constraints
\begin{align}
\label{lambda2}
0 = \e^{-2\nu(t=\phi, r=\chi)} \partial_\mu \phi \partial^\mu \phi + 1 \, , \quad
0 = \e^{-2\lambda(t=\phi, r=\chi)} \partial_\mu \chi \partial^\mu \chi - 1 \, ,
\end{align}
whose solutions are consistently given by (\ref{TSBH1}).

The constraints in Eq.~(\ref{lambda2}) render the scalar field
$\phi$ and $\chi$ non-dynamical, that is, the fluctuation of
$\phi$ and $\chi$ superimposed on the background given by (\ref{TSBH1}) do not propagate. 
In fact, by considering a perturbation of the background (\ref{TSBH1}) of the form 
\begin{align}
\label{pert1}
\phi=t + \delta \phi \, , \quad \chi=r + \delta \chi\, ,
\end{align}
and by using Eq.~(\ref{lambda2}), we find 
\begin{align}
\label{pert2}
\partial_t \delta \phi = \partial_r \delta \chi = 0\, .
\end{align}
Therefore if we impose the initial condition $\delta\phi=0$, we find $\delta\phi=0$ in the whole spacetime. 
On the other hand, if we impose the boundary condition $\delta\chi\to 0$ when $r\to \infty$,
we find $\delta\chi=0$ in the whole spacetime. 
These observations indicate that both $\phi$ and $\chi$ behave as non-dynamical or frozen degrees of freedom.

It is important to highlight that in the model described by the modified action 
$S_{\mathrm{GR} \phi\chi} + S_\lambda$, $\lambda_\phi=\lambda_\chi=0$ 
consistently appear as a solution. 
Consequently, any solution derived from the equations (\ref{gb4bD4}) and (\ref{I10}) which are based on the original action (\ref{I8})
also holds true for the modified model with the action altered according to 
$S_{\mathrm{GR} \phi\chi} \to S_{\mathrm{GR} \phi\chi} + S_\lambda$, with the additional term given by Eq.~(\ref{lambda1}). 
We now validate this statement. 

For the modified action $S_{\mathrm{GR} \phi\chi} + S_\lambda$ the original equations of motion 
given by (\ref{gb4bD4}) and (\ref{I10}) become 
\begin{align}
\label{gb4bD4mod}
0= &\, \frac{1}{2\kappa^2}\left(- R_{\mu\nu} + \frac{1}{2} g_{\mu\nu} R\right) \nonumber \\
&\, + \frac{1}{2} g_{\mu\nu} \left\{
 - \frac{1}{2} A (\phi,\chi) \partial_\rho \phi \partial^\rho \phi
 - B (\phi,\chi) \partial_\rho \phi \partial^\rho \chi
 - \frac{1}{2} C (\phi,\chi) \partial_\rho \chi \partial^\rho \chi - V (\phi,\chi)\right\} \nonumber \\
&\, + \frac{1}{2} \left\{ A (\phi,\chi) \partial_\mu \phi \partial_\nu \phi
+ B (\phi,\chi) \left( \partial_\mu \phi \partial_\nu \chi
+ \partial_\nu \phi \partial_\mu \chi \right)
+ C (\phi,\chi) \partial_\mu \chi \partial_\nu \chi \right\} \nonumber \\
&\, + \frac{1}{2}g_{\mu\nu} \left\{ \lambda_\phi \left( \e^{-2\nu(r=\chi)} \partial_\rho \phi \partial^\rho \phi + 1 \right)
+ \lambda_\chi \left( \e^{-2\lambda(r=\chi)} \partial_\rho \chi \partial^\rho \chi - 1 \right) \right\} \nonumber \\
&\, - \lambda_\phi \e^{-2\nu(r=\chi)} \partial_\mu \phi \partial_\nu \phi
 - \lambda_\chi \e^{-2\lambda(r=\chi)} \partial_\mu \chi \partial_\nu \chi + \frac{1}{2} T_{\mathrm{matter}\, \mu\nu} \, , 
\end{align}
and 
\begin{subequations}
\begin{align}
\label{I10mod_a}
0 =&\, \frac{1}{2} A_\phi \partial_\mu \phi \partial^\mu \phi
+ A \nabla^\mu \partial_\mu \phi + A_\chi \partial_\mu \phi \partial^\mu \chi
+ \left( B_\chi - \frac{1}{2} C_\phi \right)\partial_\mu \chi \partial^\mu \chi
+ B \nabla^\mu \partial_\mu \chi - V_\phi \nonumber \\
&\, - 2 \nabla^\mu \left( \lambda_\phi \e^{-2\nu(r=\chi)} \partial_\mu \phi \right) \, ,\\
 \label{I10mod_b}
0 =&\, \left( - \frac{1}{2} A_\chi + B_\phi \right) \partial_\mu \phi \partial^\mu \phi
+ B \nabla^\mu \partial_\mu \phi
+ \frac{1}{2} C_\chi \partial_\mu \chi \partial^\mu \chi + C \nabla^\mu \partial_\mu \chi
+ C_\phi \partial_\mu \phi \partial^\mu \chi - V_\chi \nonumber \\
&\, - 2 \nabla^\mu \left( \lambda_\chi \e^{-2\lambda(r=\chi)} \partial_\mu \chi \right) \, .
\end{align}
\end{subequations}
We now consider the solution of the equations (\ref{gb4bD4}) and (\ref{I10}) 
for the spherically symmetrical metric (\ref{GBiv}) with the field defined as in (\ref{TSBH1}). 
Then, the $(t,t)$ and $(r,r)$ components of Eq.~(\ref{gb4bD4mod}) give
\begin{align}
\label{lambdas}
0=\lambda_\phi = \lambda_\chi\, .
\end{align}
Other components of Eq.~ (\ref{gb4bD4mod}) are satisfied identically.
On the other hand, Eqs.~\eqref{I10mod_a} and \eqref{I10mod_b} give,
\begin{align}
\label{lambdas2}
0 = \nabla^\mu \left( \lambda_\phi \e^{-2\nu(r=\chi)} \partial_\mu \phi \right)
= \nabla^\mu \left( \lambda_\chi \e^{-2\lambda(r=\chi)} \partial_\mu \chi \right) \, .
\end{align}
Thus, if Eq.~(\ref{lambdas2}) holds true, Eq. ~(\ref{lambdas}) will also be satisfied. 
This tells us that any solution of Eqs.~(\ref{gb4bD4}) and
(\ref{I10}) is also a solution of Eqs.~(\ref{gb4bD4mod}),
(\ref{I10mod_a}), and (\ref{I10mod_b}), the latter set corresponding to the modified action
$S_{\mathrm{GR} \phi\chi} + S_\lambda$ [see Eq.~(\ref{lambda1})]
if $\lambda_\phi$ and $\lambda_\chi$ vanish accroding to Eq.~(\ref{lambdas}). 
We should note, however, that for
Eqs.~(\ref{gb4bD4mod}), (\ref{I10mod_a}), and (\ref{I10mod_b}), there
could be a solution where $\lambda_\phi$ and $\lambda_\chi$ do not vanish.

\section{Reconstruction of Models which realize any given spherically 
symmetric and static/time-dependent spacetime} 
\label{sec:Reconstruction}

\subsection{Reconstruction of model realizing given spacetime}

We now consider the problem of constructing a model which has a solution that corresponds to the geometry (\ref{GBiv}). 
For the metric~(\ref{GBiv}), the only non-vanishing connections are 
\begin{align}
\label{GBv0}
&\Gamma^t_{tt}=\dot\nu \, , \quad \Gamma^r_{tt} = \e^{-2(\lambda - \nu)}\nu' \, , \quad \Gamma^t_{tr}=\Gamma^t_{rt}=\nu'\, , \quad
\Gamma^t_{rr} = \e^{2\lambda - 2\nu}\dot\lambda \, , \quad \Gamma^r_{tr} = \Gamma^r_{rt} = \dot\lambda \, , \nonumber \\
& \Gamma^r_{rr}=\lambda'\, , \Gamma^i_{jk} = \bar{\Gamma} ^i_{jk}\, ,\quad \Gamma^r_{ij}=-\e^{-2\lambda}r \bar{g}_{ij} \, , \quad
\Gamma^i_{rj}=\Gamma^i_{jr}=\frac{1}{r} \, \delta^i_{\ j}\,,
\end{align}
where the metric $\bar{g}_{ij}$ is defined by 
$\sum_{i,j=1}^2 \bar{g}_{ij} dx^i dx^j = d\vartheta^2 + \sin^2\vartheta \, d\varphi^2$, 
$\left(x^1=\vartheta,\, x^2=\varphi\right)$ and $\bar{ \Gamma}^i_{jk}$ is the metric connection of $\bar{g}_{ij}$,
while the ``dot'' and the ``prime'' denote differentiation with respect to $t$ and $r$, respectively. 
Using the expression of the Riemann tensor $R^\lambda_{\ \mu\rho\nu}$, the Ricci tensor $R_{\mu\nu}$, and the scalar curvature $R$, 
\begin{align}
\label{Riemann}
R^\lambda_{\ \mu\rho\nu} \equiv \Gamma^\lambda_{\mu\nu,\rho} -\Gamma^\lambda_{\mu\rho,\nu} + \Gamma^\eta_{\mu\nu}\Gamma^\lambda_{\rho\eta}
 - \Gamma^\eta_{\mu\rho}\Gamma^\lambda_{\nu\eta} \, , \quad 
R_{\mu\nu} \equiv R^\rho_{\ \mu\rho\nu} \, , \quad R \equiv g^{\mu\nu} R_{\mu\nu}\, ,
\end{align}
one finds,
\begin{align}
\label{curvatures}
R_{rtrt} = &\, - \e^{2\lambda} \left\{ \ddot\lambda + \left( \dot\lambda - \dot\nu \right) \dot\lambda \right\}
+ \e^{2\nu}\left\{ \nu'' + \left(\nu' - \lambda'\right)\nu' \right\} \, ,\nonumber \\
R_{titj} =&\, r\nu' \e^{-2\lambda + 2\nu} \bar{g}_{ij} \, ,\nonumber \\
R_{rirj} =&\, \lambda' r \bar{ g}_{ij} \, ,\quad {R_{tirj}= \dot\lambda r \bar{ g}_{ij} } \, , \quad
R_{ijkl} = \left( 1 - \e^{-2\lambda}\right) r^2 \left(\bar{g}_{ik} \bar{g}_{jl} - \bar{g}_{il} \bar{g}_{jk} \right)\, ,\nonumber \\
R_{tt} =&\, - \left\{ \ddot\lambda + \left( \dot\lambda - \dot\nu \right) \dot\lambda \right\}
+ \e^{-2\lambda + 2\nu} \left\{ \nu'' + \left(\nu' - \lambda'\right)\nu' + \frac{2\nu'}{r}\right\} \, ,\nonumber \\
R_{rr} =&\, \e^{2\lambda - 2\nu} \left\{ \ddot\lambda + \left( \dot\lambda - \dot\nu \right) \dot\lambda \right\}
 - \left\{ \nu'' + \left(\nu' - \lambda'\right)\nu' \right\}
+ \frac{2 \lambda'}{r} \, ,\quad
R_{tr} =R_{rt} = \frac{2\dot\lambda}{r} \, , \nonumber \\
R_{ij} =&\, \left\{ 1 + \left\{ - 1 - r \left(\nu' - \lambda' \right)\right\} \e^{-2\lambda}\right\} \bar{g}_{ij}\ , \nonumber \\
R=& \, 2 \e^{-2 \nu} \left\{ \ddot\lambda + \left( \dot\lambda - \dot\nu \right) \dot\lambda \right\}
+ \e^{-2\lambda}\left\{ - 2\nu'' - 2\left(\nu' - \lambda'\right)\nu' - \frac{4\left(\nu' - \lambda'\right)}{r} + \frac{2\e^{2\lambda} - 2}{r^2} \right\} \, .
\end{align}
Then the $(t,t)$, $(r,r)$, $(i,j)$, and $(t,r)$ components of
(\ref{gb4bD4}) have the following forms,
\begin{subequations}
 \begin{align}
 \label{TSBH2a}
\frac{\e^{-2\lambda }}{\kappa^2} \left( \frac{2\lambda'}{r} + \frac{\e^{2\lambda} - 1}{r^2} \right)
=&\, \frac{A}{2} \e^{-2\nu} + \frac{C}{2}
 \e^{-2\lambda} + V + \rho \, ,
 \\
 \label{TSBH2b}
\frac{\e^{-2\lambda}}{\kappa^2} \left( \frac{2\nu'}{r} - \frac{\e^{2\lambda} - 1}{r^2} \right)
=&\, \frac{A}{2} \e^{-2\nu} + \frac{C}{2}
 \e^{-2\lambda} - V + p \, ,
 \\
 \label{TSBH2c}
\frac{1}{\kappa^2} \left\{ - \e^{-2 \nu} \left[ \ddot\lambda
 \right. \right. & + \left. \left. \left( \dot\lambda
 - \dot\nu \right) \dot\lambda \right]
 + \e^{-2\lambda}\left[ \frac{1}{r} \left(\nu' - \lambda' \right)
 + \nu'' + \left(\nu' - \lambda' \right) \nu' \right]\right\} \nonumber \\
=&\, \frac{A}{2} \e^{-2\nu} - \frac{C}{2} \e^{-2\lambda} -
 V + p \, , \\
 \label{TSBH2d}
\frac{2\dot\lambda}{\kappa^2 r} =&\, B \, .
\end{align}
\end{subequations}
Here we assumed that the matter is a perfect fluid and $\rho$ and $p$
are the energy density and the pressure of matter, defined by, 
\begin{align}
\label{FRk2}
T_{\mathrm{matter}\, tt}=\rho=-g_{tt}\rho\ ,\quad T_{\mathrm{matter}\, ij}=pg_{ij}\, .
\end{align}
Eqs.~\eqref{TSBH2a}-\eqref{TSBH2c} in the static limit, i.e., 
when the time-derivatives are dropped and $B=0$, describe non-rotating, spherically symmetrical compact objects. 
Their general relativistic limit is given by Tolman--Oppenheimer-Volkoff equations which can be 
recovered for $A=C=V=0$. In general the coefficients $A$, $B$, $C$,
and $V$ span a four-dimensional (4d) space of alternative theories of
gravity characterized by specific values of these parameters, i.e., by a point in this 4d-space. 
A possible strategy for finding the structure of non-rotating compact objects would be to solve 
Eqs.~\eqref{TSBH2a}-\eqref{TSBH2c} numerically for a given equation 
of state of matter, i.e., the function $p(\rho)$ at each point of the 4d parameter-space. 
We relegate this to a feature work. 
Here we will follow an alternative route, by specifying an analytical equation of
 state of matter and a profile of the star that is matched to values
 of parameters $A$, $C$, and $V$. 

To find any specific theory satisfying a prescribed equation of state and stellar profile, we can solve the set of equations 
\eqref{TSBH2a}-\eqref{TSBH2d} for the coefficients defining an alternative theory of gravity to find
\begin{align}
\label{ABCV}
A=& \frac{\e^{2\nu}}{\kappa^2} \left\{ - \e^{-2 \nu} \left[ \ddot\lambda + \left( \dot\lambda - \dot\nu \right) \dot\lambda \right]
+ \e^{-2\lambda}\left[ \frac{\nu' + \lambda'}{r} + \nu'' + \left( \nu' - \lambda' \right) \nu' + \frac{\e^{2\lambda} - 1}{r^2}\right] \right\}
 - { \e^{2\nu} \left( \rho + p \right)} \, , \nonumber \\
B=&\, \frac{2\dot\lambda}{\kappa^2 r} \, , \nonumber \\
C=&\, { \frac{\e^{2\lambda}}{\kappa^2} \left\{ \e^{-2 \nu} \left[ \ddot\lambda + \left( \dot\lambda - \dot\nu \right) \dot\lambda \right]
 - \e^{-2\lambda}\left[ - \frac{\nu' + \lambda'}{r} + \nu'' + \left( \nu' - \lambda' \right) \nu' + \frac{\e^{2\lambda} - 1}{r^2}\right] \right\} }
\, , \nonumber \\
V=& \frac{\e^{-2\lambda}}{\kappa^2} \left( \frac{\lambda' - \nu'}{r} + \frac{\e^{2\lambda} - 1}{r^2} \right) - { \frac{1}{2} \left( \rho - p \right)} \, .
\end{align}
We can now formulate a model that realizes the spacetime defined by
(\ref{GBiv}) if we find $(t,r)$-dependence of $\rho$ and $p$ and
replace $(t,r)$ in (\ref{ABCV}) with $(\phi,\chi)$.

\subsection{Other solution besides given spacetime}

We have provided a formalism to reconstruct a model that realizes a
given spacetime. Here, we examine whether there are any solutions
other than the given spacetime. For this purpose, we assume that the
given spacetime is asymptotically flat or asymptotically de Sitter
(dS) or anti-de Sitter (adS). We then demonstrate that any solution in
Einstein's gravity, with or without matter, is also a solution of the
reconstructed model.

The assumption that the spacetime is asymptotically flat, dS, or adS
when $t \rightarrow \pm \infty$ or $r \rightarrow \infty$ implies that
the static or eternal flat, dS, and adS spacetimes are also solutions automatically. 
This can be seen by considering the limit $t_0 \rightarrow \pm \infty$ after shifting the time coordinate
$t \rightarrow t_0+t$ or the limit $r_0 \rightarrow \infty$ after
shifting the radial coordinate $r \rightarrow r_0+r$. 
Although the scalar field $\phi=t_0+t$ also goes to $\pm \infty$ or $\chi=r_0+r$
goes to $\infty$ in these limits, we can redefine the scalar field to
remain finite, for example, by using the redefinitions
$\Phi={\phi}^{-1}$ or $X={\chi}^{-1}$. 
We should note that in the obtained solution describing flat, dS, or adS spacetime, 
the scalar fields $\phi$ and $\chi$ or $\Phi$ and $X$ cannot be identified with the cosmological time $t$ and 
the radial coordinate $r$ but they are constant $\Phi=X=0$ everywhere in the spacetime. 

We now consider why flat, dS, or adS spacetime is a solution to the present model. 
In the limit $\phi \rightarrow \pm \infty$ or
$\chi \rightarrow \infty$, $ A, B$, and $C$ in \eqref{ABCV} vanish,
and $V$ either vanishes or becomes a constant corresponding to the
effective cosmological constant. In terms of $\Phi$ or $X$, the
extremum (minimum or maximum) of the scalar field potential
$V\left(\Phi=\phi^{-1}, X=\chi^{-1}\right)$ is given by $\Phi=0$
and/or $X=0$. 
This is the reason why flat, dS, or adS spacetime is a solution. 
Because $\Phi$ and/or $X$ are constant, the potential plays the role of the vanishing 
or non-vanishing cosmological constant. 
We should also note that in the solution where $\Phi=0$ and/or $X=0$, 
the field equations~(\ref{I10mod_a}) and (\ref{I10mod_b}) are satisfied everywhere in the spacetime. 
Furthermore the Einstein equation in (\ref{gb4bD4mod}) 
reduces to the Einstein equation without the scalar fields $\phi$ and $\chi$, 
\begin{align}
\label{gb4bD4moddec0}
0= \frac{1}{2\kappa^2}\left(- R_{\mu\nu} + \frac{1}{2} g_{\mu\nu} R\right) + \frac{1}{2} g_{\mu\nu} \Lambda
+ \frac{1}{2} T_{\mathrm{matter}\, \mu\nu} \, .
\end{align}
Here the vanishing or non-vanishing cosmological constant is given by the potential $V$, 
$\Lambda=V\left(\Phi=0\mbox{ and/or }X=0\right)$, where the scalar field(s) is on the extreme of the potential $V$. 
Therefore, not only flat, dS, or adS spacetime, but the standard cosmological solutions like the Schwarzschild black hole, 
Kerr black hole, etc. or self-gravitating objects like standard stellar solutions including neutron stars, etc. 
in Einstein's gravity are surely also solutions in this model. 

For example, in the static case, when we neglect the contribution of matter, we can assume
$\lambda=\frac{\lambda_0}{r}+\mathcal{O}\left(r^{-2}\right)$ and
$\nu=\frac{\nu_0}{r}+$ $\mathcal{O}\left(r^{-2}\right)$ for large $r$
if the spacetime is asymptotically flat. Under these conditions, $V$
in \eqref{ABCV} behaves as:
\begin{align}
\label{Vas}
V\left(\phi \right)
= \frac{\lambda_0 + \nu_0}{\kappa^2\chi^3} + \mathcal{O}\left( \chi^{-4} \right) 
= \left( \lambda_0 + \nu_0 \right) X^3 + \mathcal{O}\left( X^4 \right)\, .
\end{align}
Therefore $X=0$ is surely an extremum of the potential $V$. 
When $X=0$, the field equations~(\ref{I10mod_a}) and (\ref{I10mod_b}) are trivially satisfied. 
Therefore $X=0$ is an exact solution of the field equations~(\ref{I10mod_a}) and (\ref{I10mod_b}). 
We should note that in the solution, $\chi=X^{-1}$ and $\phi=\Phi^{-1}$ are not identified with $r$ and $t$. 
Therefore the solution is not included in the class of the solution satisfying Eq.~(\ref{TSBH1}). 
Then we can consider the solution where the scalar fields stay at the extremum. 
For the solution, the scalar fields decouple from gravity 
because $A$, $B$, and $C$ in (\ref{ABCV}) vanish and $V$ also vanishes
or becomes a constant corresponding to the effective cosmological constant and the Einstein equation in (\ref{gb4bD4mod}) 
reduces to the Einstein equation (\ref{gb4bD4moddec0}) without the scalar fields $\phi$ and $\chi$. 
Therefore when the energy-momentum tensor vanishes
($T_{\mathrm{matter}\, \mu\nu}=0$) the Schwarzschild solution is a solution of this model. 
Because the model in this paper reduces to the standard Einstein gravity with/without matter in
(\ref{gb4bD4moddec0}), any solution in the standard Einstein gravity with/without matter is also a solution of the present model.

We should also note that, since the model does not explicitly depend
on any specific coordinates, an object like the extremely compact star
described in this paper could form in any location. This implies that
there should be solutions where multiple stars exist, provided the
distances between them are large enough to neglect non-linear
interactions. Therefore, although it is impossible to identify all
solutions within the model, we can conclude that the model
accommodates a rich variety of solutions.

\section{Model realizing spacetime where the radius of the stellar object is smaller than the Schwarzschild radius}
\label{sec:StaticStars}

In the following, we assume the spacetime is static, that is,
$\nu=\nu(r)$ and $\lambda=\lambda(r)$, and drop time-derivatives.
Then the set of equations (\ref{ABCV}) reduces to the following forms, 
\begin{align}
\label{ABCVstatic}
A=& \frac{\e^{2\nu-2\lambda}}{\kappa^2} \left( \frac{\nu' + \lambda'}{r} + \nu'' + \left( \nu' - \lambda' \right) \nu' + \frac{\e^{2\lambda} - 1}{r^2}\right) 
 - { \e^{2\nu} \left( \rho + p \right)} \, , \nonumber \\
B=&\, 0 \, , \nonumber \\
C=&\, { - \frac{1}{\kappa^2} \left( - \frac{\nu' + \lambda'}{r} + \nu'' + \left( \nu' - \lambda' \right) \nu' + \frac{\e^{2\lambda} - 1}{r^2}\right)} \, , \nonumber \\
V=& \frac{\e^{-2\lambda}}{\kappa^2} \left( \frac{\lambda' - \nu'}{r} + \frac{\e^{2\lambda} - 1}{r^2} \right) - { \frac{1}{2} \left( \rho - p \right)} \, .
\end{align}
We assume that the energy density $\rho$ and the pressure $p$ of
matter satisfies the cold (zero-temperature) equation of state, $p=p\left(\rho\right)$. 
Because we are considering static spacetime, $\rho$ and $p$ depend on $r$, only. 
Therefore the energy density $\rho$ and the pressure $p$ satisfy the following conservation law, 
which can be obtained from the equations in (\ref{TSBH2a}), (\ref{TSBH2b}), (\ref{TSBH2c}), and (\ref{TSBH2d}), 
\begin{equation}
\label{FRN2}
0 = \nabla^\mu T_{\mu r} =\nu' \left( \rho + p \right) + p' \, .
\end{equation}
Other components of the conservation law are trivially satisfied.
If the equation of state $\rho=\rho(p)$ is given, Eq.~(\ref{FRN2}) can be integrated as
\begin{equation}
\label{FRN3}
\nu = - \int^r dr \frac{\frac{dp}{dr}}{\rho + p}
= - \int^{p(r)}\frac{dp}{\rho(p) + p} \, .
\end{equation}

The equation of state of compact stars, like neutron stars, often can
be approximated by a polytrope. 
We consider two options: 
\begin{enumerate}
\item Energy-polytrope
\begin{equation}
\label{polytrope}
p = K \rho^{1 + \frac{1}{n}}\,,
\end{equation}
with constants $K$ and $n$.
It is known that for the neutron stars, $n$ could take the value in
the range $0.5\leq n \leq 1$.
\item Mass-polytrope
\begin{equation}
\label{MassPolytropicEOS}
\rho = \rho_{m} + N p \, ,\qquad \qquad p = K_m \rho_m^{1+\frac{1}{n_{m}}} \, ,
\end{equation}
where $\rho_{m}$ is rest mass energy density and $K_{m}$ and $N$ are constants.
\end{enumerate}

It is convenient to rewrite Eq.~\eqref{polytrope} as
\begin{align}
\label{polytrope2}
\rho = \tilde K p^{({1 + \frac{1}{\tilde n}})}\, , \quad
\tilde K \equiv K^{-\frac{1}{1+\frac{1}{n}}} \, , \quad
\tilde n \equiv \frac{1}{\frac{1}{1+\frac{1}{n}} - 1}
= - 1 - n \, .
\end{align}
In this case, Eq.~(\ref{FRN3}) takes the following form,
\begin{align}
\label{FRN3p1B}
\nu = - \int^{p(r)}\frac{dp}{\tilde K p^{1 + \frac{1}{\tilde n}} + p}
= \frac{c}{2} + \tilde n \ln \left(1+{\tilde K}^{-1}p^{-\frac{1}{\tilde n}} \right)
= \frac{c}{2} - \left(1+n\right) \ln \left(1+ K \rho^\frac{1}{n} \right) \, ,
\end{align}
where $c$ is a constant of the integration.
Similarly in case of mass-polytrope (\ref{MassPolytropicEOS}), we obtain
\begin{align}
\label{masspolytope}
\nu = \frac{\tilde c}{2} + \ln \left( 1 - K_m \rho^\frac{1}{n_m}\right) \, ,
\end{align}
where $\tilde c$ is another constant of the integration.

Given one of the aforementioned equations of state, let us consider an example case 
where we assume the following profile for
$\rho = \rho(r)$ and $\lambda = \lambda(r)$ with 
\begin{align}
\label{anz1}
\rho=\left\{ \begin{array}{cc}
\rho_c \left( 1 - \frac{r^2}{{R_s}^2} \right) & \ \mbox{when}\ r<R_s \\
0 & \ \ \mbox{when}\ r>R_s
\end{array} \right. \, , \quad
\e^{-2\lambda} = 1 - \frac{2Mr^2}{r^3 + {r_0}^3} \, ,
\end{align}
where $r_0$ is a constant, $\rho_c$ is the energy density at the centre of the star, and $R_s$ is also 
a constant corresponding to the radius of the surface of the compact
star, which corresponds to the ADM mass. 
When $r\to \infty$, $\e^{-2\lambda}$ behaves as $\e^{-2\lambda} \sim 1 - \frac{2M}r$ and therefore $M$ can be
regarded as the mass of the compact star.

In general, due to the contribution of the scalar fields, the ADM mass
$M$ is different from the mass $M_\mathrm{\star}$ corresponding to
ordinary matter of the compact star given by 
\begin{align}
\label{MRs}
M_\mathrm{\star} =4\pi \rho_c \int_0^r \psi^2 \rho(\psi) d\psi= 4\pi \rho_c \int_0^r d\psi \psi^2 \left( 1 - \frac{\psi^2}{{R_s}^2} \right)
= \frac{4\pi \rho_c r^3}{15}\left( 5 - \frac{3r^2}{{R_s}^2} \right)\,.
\end{align}
We also note that we need to choose $r_0$ large enough so that $\e^{2\lambda}$ is positive. 
For $\e^{-2\lambda}$ in \eqref{anz1} to be positive, it is necessary that 
\begin{align}
\label{cond2}
\frac{{r_0}^3}{M^3}>\frac{32}{27} \, .
\end{align}
We should also note that when $r\to 0$, $\e^{-2\lambda}$ behaves as
$\e^{-2\lambda(r)} \sim 1 - \frac{2Mr^2}{r_0^3}$. 
Therefore $\lambda'(r)$ vanishes at the center, i.e., $\lambda'(r=0)=0$, and
therefore, there is no conical singularity.

As an example, let us adopt the energy-polytope as the equation of
state and select $n=1$ for similicity. 
Then Eq.~(\ref{FRN3p1B}) gives
\begin{align}
\label{FRN3p1BC1}
\e^{2\nu} = \frac{\e^c}{ \left( 1 + K \rho_c \left( 1 - \frac{r^2}{{R_s}^2} \right) \right)^4} \, ,
\end{align}
and, consequently, 
\begin{align}
\label{FRN3p1BC2}
\left( \e^{2\nu} \right)' = \frac{8 \e^c K \rho_c r}{{R_s}^2\left( 1 + K \rho_c \left( 1 - \frac{r^2}{{R_s}^2} \right) \right)^5} \, .
\end{align}
Outside the star, we assume $\e^{2\nu(r)}=\e^{-2\lambda(r)}$ in (\ref{anz1}) and therefore
\begin{align}
\label{arprimeout}
\left(\e^{2\nu}\right)' = \frac{2Mr\left( r^3 - {r_0}^3\right)}{\left( r^3 + {r_0}^3\right)^2} \, .
\end{align}
Because $\e^{2\nu(r)}$ and $\left(\e^{2\nu(r)}\right)'$ should be continuous at the surface $r=R_s$. we obtain
\begin{align}
\label{matching}
\e^c = 1 - \frac{2M{R_s}^2}{{R_s}^3 + {r_0}^3} \, , \quad
\frac{8 \e^c K \rho_c}{R_s} = \frac{2M{R_s}\left( {R_s}^3 - {r_0}^3\right)}{\left( {R_s}^3 + {r_0}^3\right)^2} \, .
\end{align}
After eliminating $\e^c$ from the two equations in \eqref{matching} we obtain
\begin{align}
\label{r0eq}
0= \left({r_0}^3\right)^2 + \left( 2 {R_s}^3 - 2M{R_s}^2 + \frac{M{R_s}^2}{4K \rho_c} \right){r_0}^3 + {R_s}^6 - 2M{R_s}^5 - \frac{M{R_s}^5}{4K \rho_c} \, .
\end{align}
Because $r_0$ should be positive, we need to impose the condition
\begin{align}
\label{cond}
& R_s - 2M - \frac{M}{4K \rho_c} <0 \nonumber \\
& \mbox{or} \nonumber \\
& 2 R_s - 2M + \frac{M}{4K \rho_c} <0 \quad \mbox{and} \quad
R_s - 2M - \frac{M}{4K \rho_c} > 0 \, .
\end{align}
Because $K$ and $\rho_c$ are positive parameters, the second condition cannot be satisfied. 
It is convenient to rewrite the first condition as 
\begin{align}
\label{condB}
R_s < 2M + \frac{M}{4K \rho_c} \, ,
\end{align}
where $R_\mathrm{Schwarzschild}=2M$ in the units we have adopted corresponds to the Schwarzschild radius. 
An interesting feature of Eq.~(\ref{cond2}) is
that it is satisfied even in the case where the radius $R_s$ of the neutron star is smaller 
than the Schwarzschild radius $R_\mathrm{Schwarzschild}$. 
Another issue to examine is whether the condition (\ref{cond2}) can be satisfied. 
In general, the condition (\ref{cond2}) may not be satisfied
even if Eq.~(\ref{condB}) is satisfied. 
We rewrite Eq.~(\ref{r0eq}) as follows, 
\begin{align}
\label{r0eqB}
0= \mathcal{F}\left(\frac{{r_0}^3}{{R_s}^3}\right) \equiv \left(\frac{{r_0}^3}{{R_s}^3}\right)^2 
+ \left( 2 - \frac{2M}{R_s} + \frac{M}{R_s}\frac{1}{4K \rho_c} \right)\frac{{r_0}^3}{{R_s}^3} 
+ 1 - \frac{2M}{R_s} - \frac{M}{R_s}\frac{1}{4K \rho_c} \, .
\end{align}
We now note that $\mathcal{F}(0)<0$ if the first condition in (\ref{cond}) is satisfied. 
If further $\mathcal{F}\left( \frac{32}{27} \right)>0$, there is a solution satisfying (\ref{cond2}). 
The condition in (\ref{condB}) can be rewritten as follows, 
\begin{align}
\label{condB2}
\epsilon \equiv \frac{2M}{R_s} + \frac{M}{R_s}\frac{1}{4K\rho_c} -1 > 0 \, .
\end{align}
By using (\ref{condB2}), we find 
\begin{align}
\label{condB3}
\mathcal{F}\left( \frac{32}{27} \right) = \left( \frac{32}{27} \right)^2 + \left( 3 - \frac{4M}{R_s} + \epsilon \right) \frac{32}{27} - \epsilon \, .
\end{align}
For $\mathcal{F}\left( \frac{32}{27} \right)$ to be positive, we determine that
\begin{align}
\label{condB4}
\frac{2M}{R_s}< \frac{1}{2} \left( \frac{113}{27} - \frac{59}{32} \epsilon\right) \, .
\end{align}
Therefore as long as 
\begin{align}
\label{condB5}
0< \epsilon < \frac{3616}{1593}\, ,
\end{align}
the neutron stars radius $R_s <R_\mathrm{Schwarzschild}=2M$. Eq.~(\ref{condB5}) gives a constraint on $K\rho_c$, 
\begin{align}
\label{condB6}
\frac{1}{4K\rho_c} < - 2 + \frac{5209 R_s}{1593 M} \, .
\end{align}
For example, if assume that $R_s=M<2M=R_\mathrm{Schwarzschild}$, we find that
$\frac{1}{4K\rho_c} < \frac{2023}{1593}$ and, therefore,
$\rho_c > \frac{1}{4K} \frac {1593}{2023}$. 
Thus obtaining small radii requires large central densities which will result in large
masses of extremely compact neutron stars. 
Thus, in principle, for sufficiently large central masses we find that the radius $R_s$ 
of a neutron star can be smaller than the Schwarzschild radius $R_\mathrm{Schwarzschild}$.

Finally, we consider the case where Eq.~(\ref{cond2}) is not satisfied. 
In this particular case, $\e^{-2\lambda}$ in (\ref{anz1}) has
two zero points $\e^{-2\lambda(r=r_\pm)}=0$, $r=r_\pm>0$ assuming $r_+>r_-$. 
On the other hand, according to Eq.~(\ref{FRN3p1BC1})
$\e^{2\nu}$ does not vanish inside the star although
$\e^{2\nu}=\e^{-2\lambda}$ outside the star. Then there could be three cases:
\begin{enumerate}
\item $R_s>r_+$: in the region $r_-<r<r_+$, two directions become time-like, resulting in an inconsistency.
\item $r_+>R_s>r_-$: The region $r_-<r<r_+$ can be regarded as the
 region inside the black hole and therefore the stellar object is collapsed. 
\item $R_s<r_-$: corresponds to the case where there is a black hole with two horizons. 
Given the standard signature within the region $r<r_-$, a stellar object can indeed exist there. 
Consequently, this black hole has regular characteristics. 
It is akin to the Reissner-Nordstr{\"o}m black hole featuring a finite-sized electric
 charge along with mass situated at its centre.
\end{enumerate}
The last case may solve the problem of the black hole singularity. 
Recently there has appeared a scenario where the signature inside the horizon becomes Euclidean and the singularity can be avoided~\cite{Capozziello:2024ucm}.

\section{Formation of the extremely compact star}
\label{sec:Formation}

We may consider how the stellar object may be formed by the collapse of the matter. 
If the matter couples with the two scalar fields, the cloud of the scalar fields may be also formed by the collapse. 
Even if the scalar fields do not couple with matter, they may also collapse due to gravity. 
In the latter case, because the scalar fields satisfy the conservation law independent of that of the matter, 
there should exist the scalar fields even uniform from the beginning.

In the following, we consider the latter case and we show that a neutron star with a radius larger 
than the Schwarzshild radius can collapse into a stellar object with a radius smaller than the Schwarzschild one. 
We consider the time-dependent metric in (\ref{GBiv}). 
We should note that the equations in (\ref{ABCV}) can
be used even if we consider the dynamical spacetime corresponding to the collapse. 
The matter should collapse so that the conservation law
$0=\nabla^\mu T_{\mathrm{matter}\, \mu\nu}$ is satisfied. 
Now we find
\begin{align}
\label{emt2}
T_{\mathrm{matter}\, tt} = -\rho (t,r) g_{tt} = \e^{2\nu (t,r)} \rho (t,r)\, , \quad
T_{\mathrm{matter}\, rr} = p (t,r)g_{ij} = \e^{2\lambda (t,r)} p (t,r)\, .
\end{align}
Then the conservation law is given by 
\begin{align}
\label{cnsvr1}
0 = - \dot\rho - \dot\lambda \left( \rho + p \right) \, , \quad 
0 = \nu' \left( \rho + p \right) + p' \, .
\end{align}
In a way similar to obtain (\ref{FRN3}), we find 
\begin{align}
\label{FRN3B}
\nu (t,r) = - \int^{p(t,r)}\frac{dp}{\rho(p) + p} + \nu_0 (t) \, , \quad 
\lambda (t,r) = - \int^{\rho (t,r)}\frac{d\rho}{\rho + p(\rho)} + \lambda_0 (r) \, . 
\end{align}
Here $\nu_0(t)$ and $\lambda_0(r)$ are arbitrary functions of $t$ and $r$, respectively. 

We may consider the energy-polytrope in (\ref{polytrope}), the mass-polytrope 
in (\ref{MassPolytropicEOS}), or a more general one as
the equation of state by expressing any phase transitions. 
We may also consider the time-dependent profile of matter as a generalization of (\ref{anz1d}),
\begin{align}
\label{anz1d}
\rho (t,r) =\left\{ \begin{array}{cc}
\rho_c (t) \left( 1 - \frac{r^2}{{R_s(t) }^2} \right) & \ \mbox{when}\ r<R_s(t) \\
0 & \ \ \mbox{when}\ r>R_s(t)
\end{array} \right. \, .
\end{align}
Inside the stellar object, $\nu$ and $\lambda$ are given by (\ref{FRN3B}). 
Outside the stellar object, as a time-dependent extension of 
\begin{align}
\label{profilesnulambda}
\e^{2\nu(t,r)} = 1 - \frac{2Mr^2}{r^3 + {r_0^{(\nu)}(t)}^3} \, , \quad 
\e^{-2\lambda(t,r)} = 1 - \frac{2Mr^2}{r^3 + {r_0^{(\lambda)}(t)}^3} \quad \mbox{when}\ r>R_s(t)\, .
\end{align}
As the boundary conditions, we may impose the continuities of $\nu$,
$\nu'$, $\lambda$, and $\lambda'$ at the surface $r=R_s(t)$. 
The boundary conditions could be satisfied by adjusting four functions
$\nu_0 (t)$, $\rho_c (t)$, $r_0^{(\nu)}(t)$, and $r_0^{(\lambda)}(t)$.

As an initial condition, we may choose $R_s(t)$ so that the surface
radius $R_s(t)$ is much larger than the Schwarzschild radius $R_\mathrm{Schwarzschild}=2M$ but
we may also choose the function $R_s(t)$ so that the radius $R_s(t)$
becomes smaller than the Schwarzschild radius. 
In the scenario, the neutron star does not collapse into a black hole but it becomes 
a stellar object without a horizon but the radius is smaller than the Schwarzschild radius given by the ADM mass.

\section{Summary and discussion}
\label{sec:Conclusions}

In this paper, we have constructed a model that admits a spherically
symmetric solution for a stellar object, for example a neutron star,
with $R_s< R_\mathrm{Schwarzschild}$, i.e., a radius that is smaller than the
Schwarzschild radius. 
The model we employed is given by Einstein's gravity coupled to two scalar fields \cite{Nojiri:2020blr}. 
As in the papers \cite{Nojiri:2023dvf, Nojiri:2023zlp, Elizalde:2023rds}, the ghosts in the model of
\cite{Nojiri:2020blr} are eliminated by the constraints similar to the constraint that appeared 
in the mimetic gravity theoretical framework~\cite{Chamseddine:2013kea}.

To give a concrete example, we approximated the equation of state of neutron star matter 
by an energy-polytrope (\ref{polytrope}) with $n=1$. 
The density profile of the neutron star is given by Eq.~(\ref{anz1}).
The surface of the star is defined conventionally, as the surface $r=R_s$ 
on which the energy density $\rho$ and the pressure
$p$ (related in our case by the polytropic equation of state
(\ref{polytrope}) with $n=1$) vanish. 
For the metric given by Eq.~(\ref{GBiv}), we have chosen $\lambda$ as in (\ref{anz1}). 
The parameter $\nu$ is given by (\ref{FRN3p1BC1}) inside the star and it
is assumed that $\nu=-\lambda$ outside the star. 
Under these assumptions, we showed that if the parameters of our model satisfy the
condition (\ref{condB6}), there exists a solution where the radius
$R_s<R_\mathrm{Schwarzschild}$, i.e., the radius of the neutron star is smaller than the
Schwarzschild radius. 
The obtained quantities and conditions are
summarized in TABLE~\ref{Table1}.


Traditionally, one defines a model of gravity first and finds the spacetime structure of a compact object which emerges as a 
solution of the gravity model. 
In this paper, however, we did the inverse, that is, we first gave the spacetime geometry of a compact object 
defined by a specific equation of state and density profile, and after that, we constructed a model of gravity where the given 
spacetime is a solution. 
This shows that the model realizing the given spacetime surely exists, which is the main claim of this paper. 
Clearly, the gravity model space defined by $A$, $B$, $C$, and $V$ coefficients should have many kinds of other solutions 
and we have shown that any solution of Einstein's gravity with/without matter is a solution of this model. 
In any case, the set of standard partial differential equations \eqref{TSBH2a}-\eqref{TSBH2d} which includes 
non-zero $A$, $B$, $C$, and $V$ can be utilized to define a gravity theory according to \eqref{ABCV} if any initial profile of 
the matter distribution and its equation of state are given.


\begin{table}
\caption{Obtained Quantities and Conditions}
\label{Table1}
\centering
\begin{tabular}{|c|c|}
\hline
Equation of state of matter & Energy-polytrope with $n=1$, $p = K \rho^2$ \\
\hline
Density profile of neutron star & $\rho=\left\{ \begin{array}{cc}
\rho_c \left( 1 - \frac{r^2}{{R_s}^2} \right) & \ \mbox{when}\ r<R_s \\
0 & \ \ \mbox{when}\ r>R_s
\end{array} \right.$ \\
\hline
Components of metric & $\e^{-2\lambda}=1 - \frac{2Mr^2}{r^3 + {r_0}^3}$ \\
& $\e^{2\nu} =\left\{ \begin{array}{cc}
\frac{\e^c}{ \left( 1 + K \rho_c \left( 1 - \frac{r^2}{{R_s}^2} \right) \right)^4} & \ \mbox{when}\ r<R_s \\
1 - \frac{2Mr^2}{r^3 + {r_0}^3} & \ \ \mbox{when}\ r>R_s
\end{array} \right.$ \\
\hline
ADM mass & $M$ \\
\hline
Mass of ordinary matter only & $M_{\star} = \frac{4\pi \rho_c r^3}{15}\left( 5 - \frac{3r^2}{{R_s}^2} \right)$ \\
\hline
Condition for the existence of a regular solution &
$\frac{{r_0}^3}{M^3}>\frac{32}{27}$, $R_s < 2M + \frac{M}{4K \rho_c}$ \\
\hline
\begin{minipage}{7cm}
\ \\
The condition that the radius of a neutron star \\ 
can be smaller than the Schwarzschild radius \\
\ 
\end{minipage}
& $\frac{1}{4K\rho_c} < - 2 + \frac{5209 R_s}{1593 M}$ \\
\hline
Condition for the existence of a regular black hole &
$\frac{{r_0}^3}{M^3}<\frac{32}{27}$, $r_->R_s$ \\ 
\hline
\end{tabular}
\end{table}

We now provide some speculations.

\begin{itemize}
\item If the two scalar fields operate independently from electromagnetic fields, as commonly assumed, 
the radius of the stellar object might become observable through electromagnetic radiation. 
 From an observational perspective, accounting for the photon sphere becomes essential because the radius is typically 
$1.5$ times larger than the horizon radius in Einstein's gravity.
Determining the photon sphere's radius might not pose excessive difficulty because, in this paper's model, the geometry outside 
the neutron star resembles that of the Hayward black hole \cite{Hayward:2005gi}, lacking horizons 
(see \cite{Heydari-Fard:2022jdu, Guo:2022ghl} for examples). 
However, in more general solutions, the gravitational effects induced by the scalar fields could potentially complicate the photon's orbit. 

\item It could be interesting to consider the scenario of a merger
 involving an extremely small neutron star either in a neutron star
 binary merger or in a merger of one neutron star and a black hole.
 A radius that is smaller than the Schwarzschild radius, along with the
 presence of the scalar field cloud, could potentially influence the
 signal emitted during such mergers. A specific scenario is the
 measurement of tidal deformability of merging object(s) through the
 gravitational waves. This may allow the identification of compact
 stars with very small radii, as such radii would imply tidal
 deformabilities much smaller than those of ordinary neutron stars.
 Such objects will also possess anomalously small moments of inertia.
 It remains to clarify the stability criteria of such objects against
 oscillation modes, for example, $f$-modes by which ordinary neutron
 stars become unstable as the central density of the object is
 increased.

\item A black hole has the Hawking temperature, which is very low
 {compared to the characteristic surface temperature of a mature
 neutron star, which is of the order of $10^6$~K. For example, the
 group of seven nearby X-ray dim isolated neutron stars RX
 J1856.5-3754, 0720.4-3125, 1605.3+3249, 1308.6+2127, 2143.0+0654,
 0806.4-4123 and 0806.4-4123 emits X-ray radiation characterised by
 a near-blackbody continuum in soft X-rays. Their near-thermal
 emission suggests emission from the surface with temperatures
 ranging from approximately 50 to 100 eV ~\cite{Haberl:2006xe}.}
 { Another example is PSR~J0437-4715 which has the
 bulk surface temperature $(2.3 \pm 0.1 ) \times 10^5\, \mathrm{K}$
 \cite{Gonzalez-Caniulef:2019wzi}. } {We anticipate that the
 stellar objects, studied in this work, should have comparable surface
 temperatures if they are composed of ordinary nuclear
 matter and thus can be distinguished from black
 holes once their surface temperatures are measured.}

\item We may also consider the stellar object whose radius is smaller
 than the Schwarzschild radius in other modified gravity theories.
 In the cases of the $F(R)$ gravity and the
 scalar--Einstein--Gauss-Bonnet gravity, there appears scalar mode.
 If there are clouds of the scalar mode around the neutron star, the
 scalar field gives additional contributions to the ADM mass and
 therefore the corresponding Schwarzshild radius given by the ADM
 mass can be larger than the radius of the neutron star similar to
 the case considered in this paper.

\item As a similar object, we may consider the wormhole where a neutron star could hide the throat 
by choosing the radius of the neutron star to be larger than the radius of the throat. 

\end{itemize}

\section*{Acknowledgements}

This work was partially supported by MICINN (Spain), project
PID2019-104397GB-I00 and by the program Unidad de Excelencia Maria de
Maeztu CEX2020-001058-M, Spain (S.D.O). S.N. was partly supported by
MdM Core visiting professorship at ICE-CSIC,
Barcelona. A. S. gratefully acknowledges the Polish NCN Grant
No. 2020/37/B/ST9/01937 at Wroc\l{}aw University.



\end{document}